\documentclass[twocolumn,superscriptaddress,preprintnumbers,amsmath,amssymb,pra]{revtex4-2}

\usepackage{graphicx}
\usepackage{dcolumn}
\usepackage{bm}
\usepackage{ifpdf}
\usepackage{epstopdf}
\usepackage{slashed}
\usepackage{amsfonts}
\usepackage{mathrsfs}
\usepackage{amssymb}
\usepackage{multirow}
\usepackage{CJK}
\usepackage{xcolor}
\usepackage{textcomp}
\usepackage{csquotes}

\def\lesssim{\ \raise.3ex\hbox{$<$}\kern-0.8em\lower.7ex\hbox{$\sim$}\ }
\def\gesim{\ \raise.3ex\hbox{$>$}\kern-0.8em\lower.7ex\hbox{$\sim$}\ }

\newcommand{\red}[1]{{{#1}}}

\begin{document}
\begin{CJK}{UTF8}{ipxm}
\preprint{RIKEN-iTHEMS-Report-22}

\title{Stability against three-body clustering in one-dimensional spinless $p$-wave fermions}

\author{Yixin Guo (郭一昕)}
\email{guoyixin1997@g.ecc.u-tokyo.ac.jp}
\affiliation{Department of Physics, Graduate School of Science, The University of Tokyo, Tokyo 113-0033, Japan}
\affiliation{RIKEN iTHEMS, Wako 351-0198, Japan}

\author{Hiroyuki Tajima (田島裕之)}
\email{htajima@g.ecc.u-tokyo.ac.jp}
\affiliation{Department of Physics, Graduate School of Science, The University of Tokyo, Tokyo 113-0033, Japan}

\date{\today}

\begin{abstract}
We theoretically investigate in-medium two- and three-body correlations in one-dimensional spinless fermions with attractive two-body $p$-wave interaction. 
By investigating the variational problem of two- and three-body states above the Fermi sea, we elucidate the fate of the in-medium two- and three-body cluster states.
The one-dimensional system with the strong $p$-wave interaction is found to be stable against the formation of three-body clusters even in the presence of the Fermi sea, in contrast to higher-dimensional systems that suffer the strong three-body loss associated with the trimer formation.
Our results indicate that the weak two-body coupling side is more sensitive to the residual three-body interaction than the strong-coupling side.
By including the dimensionless three-body coupling such that the universality associated with the scattering length is maintained, we find that an in-medium three-body state similar to a squeezed Cooper triple appears.

\end{abstract}

\maketitle

\section{Introduction}\label{sec:I}

The study of unconventional superconductors and superfluids is one of the most exciting issues in modern physics~\cite{Sigrist1991RevModPhys.63.239}. 
In particular, the importance of the $p$-wave pairing state has been widely recognized in the field of condensed-matter physics as well as nuclear physics, such as the $^3$He superfluid~\cite{leggett2006quantum} and the $^3$P$_2$ neutron superfluid~\cite{Dean2003RevModPhys.75.607}.

An ultracold Fermi gas near the $p$-wave Feshbach resonance is one of the promising candidates to systematically investigate the role of strong $p$-wave interaction in unconventional superfluid states~\cite{Gurarie2007AnnPhys.322.2} due to its tunable interaction~\cite{Chin2010Rev.Mod.Phys.82.1225--1286}.
In this regard, the realization of a $p$-wave superfluid Fermi gas is a long-standing concern in cold-atom physics.
A cold-atom quantum simulation of $p$-wave superfluids will make an important breakthrough in our understanding of topological superconductors~\cite{Sato2017}. 
However, in three dimensions, the $p$-wave Fermi superfluid state is unstable against three-body clustering~\cite{Levinsen2007PhysRevLett.99.210402},
which leads to the three-body recombination accompanying strong particle loss.
Such a three-body loss near the $p$-wave Feshbach resonance has been measured in experiments~\cite{Waseem2018PhysRevA.98.020702,Yoshida2018PhysRevLett.120.133401}.
Also, the three-body loss has been theoretically examined in connection with few-body physics in three dimensions~\cite{Suno2003PhysRevLett.90.053202,Jona2008PhysRevA.77.043611,Schmidt2020PhysRevA.101.062702,zhu2022three}.

In contrast, the suppression of three-body loss in a one-dimensional $p$-wave system has been theoretically predicted~\cite{Zhou2017PhysRevA.96.030701}.
In this context, the three-body loss in one-dimensional $p$-wave fermions was experimentally studied recently~\cite{Chang2020PhysRevLett.125.263402,marcum2020suppression}.
The one-dimensional $p$-wave superfluid is also relevant for the Majorana edge mode~\cite{Kitaev2001}.
Moreover, a Bose-Fermi duality is also of great interest in one-dimensional $p$-wave Fermi systems~\cite{girardeau1960relationship,Cheon1999PhysRevLett.82.2536,Yukalov2005,Sekino2018PhysRevA.97.013621,Valiente2020PhysRevA.102.053304,Valiente2021PhysRevA.103.L021302,Sekino2021PhysRevA.103.043307}.
Thanks to such a fascinating property in this system, the bulk viscosity has attracted attention~\cite{tanaka2022bulk,maki2022viscous}.
The one-dimensional $p$-wave contact has also been investigated by using the thermodynamic Bethe ansatz~\cite{Yin2018PhysRevA.98.023605} and virial expansion~\cite{Maki2021PhysRevA.104.063314}.

To investigate the stability against three-body clustering in quantum many-body systems,
we need to consider the in-medium three-body problem.
For such a purpose, the generalized Cooper problem has been applied to cluster states consisting of more than two particles, such as Cooper triples~\cite{Niemann2012Phys.Rev.A86.013628,Kirk2017Phys.Rev.A96.053614,Tajima2021Phys.Rev.A104.053328,Tajima2022Phys.Rev.Research4.L012021} and even Cooper quartets~\cite{Roepke1998Phys.Rev.Lett.80.3177--3180,Sandulescu2012Phys.Rev.C85.061303,Guo2022Phys.Rev.C105.024317,Guo2022Phys.Rev.Research4.023152}.
The investigation of the fate of such higher-order clustering is also a stimulating topic in various fields.
These approaches are useful for further understanding the many-body ground state. Indeed, the Cooper pair problem has been considered to elucidate the pairing mechanism in unconventional superconductors~\cite{Mattis1986RevModPhys.58.361,Lages2001PhysRevB.64.094502,Martikainen2008PhysRevA.78.035602,sanayei2021cooper}.
On the other hand, the corresponding study of one-dimensional spinless $p$-wave fermionic systems has not been strictly performed so far.

In this paper, we theoretically study in-medium two- and three-body states in a one-dimensional Fermi gas with the resonant $p$-wave interaction using the variational method based on the generalized Cooper problem. 
We show that a $p$-wave Cooper pair appears even in the weak-coupling regime and undergoes the crossover towards the molecular state like the BCS-BEC crossover in a three-dimensional Fermi gas with strong $s$-wave interaction~\cite{Strinati2018Phys.Rep.738.1--76,Ohashi2020Prog.Part.Nucl.Phys.111.103739}.
By solving the in-medium three-body equation derived from the variational principle, we show the absence of the stable in-medium three-body cluster state (such as a Cooper triple and in-medium trimer) in the present system without any additional interactions such as three-body attraction.
The fermion-dimer repulsion, which suppresses the in-medium three-body clustering, is found to be always present, although it is weakened by the in-medium effect in the crossover regime.
Accordingly, we also show the emergence of the in-medium three-body state is similar to the squeezed Cooper triple~\cite{Tajima2021Phys.Rev.A104.053328,Tajima2022Phys.Rev.Research4.L012021} in the presence of the residual three-body interaction proposed in Ref.~\cite{Sekino2021PhysRevA.103.043307}.

This paper is organized as follows.
In Sec.~\ref{sec:II}, we introduce
the Hamiltonian for one-dimensional spinless fermions with attractive $p$-wave two-body interaction. 
In Sec.~\ref{sec:III},
we derive the Cooper-pair problem in the one-dimensional spinless $p$-wave fermionic system.
In addition, the similarity between the present Cooper problem and mean-field theory is discussed in the Appendix.
In Sec.~\ref{sec:IV},
we show
the in-medium three-body equation obtained from the variational approach and the effective fermion-dimer repulsion.
The three-body bound state is found to be absent in one-dimensional spinless $p$-wave fermions correspondingly.
Finally, a summary and perspectives are given in Sec.~\ref{sec:V}.
In the following, we take $\hbar=c=k_{\rm B}=1$.
The system size is taken to be unit.

\section{Hamiltonian}\label{sec:II}

In this paper, we consider one-dimensional spinless fermions with attractive $p$-wave interaction.
The Hamiltonian of such a system can be given as
\begin{align}
    H=\,&K+V,\\
    K=\,&\sum_{k}\xi_{k}c_{k}^\dag c_{k},\\
    V=\,&\frac{U_2}{2}\sum_{k_1,k_2,k_3,k_4}
    \left(\frac{k_1-k_2}{2}\right)
    \left(\frac{k_3-k_4}{2}\right)\nonumber\\
    &\times c_{k_1}^\dag c_{k_2}^\dag 
    c_{k_4} c_{k_3}
    \delta_{k_1+k_2,k_3+k_4},
\end{align}
where $\xi_{k}=k^2/(2m)-\mu$ in the kinetic term $K$ is the single-particle energy with momentum $k$, atomic mass $m$, and chemical potential $\mu$.
In the generalized Cooper problems, we take $\mu=E_{\rm F}$, where $E_{\rm F}$ is the Fermi energy.
$V$ represents the short-range $p$-wave two-body interaction with coupling constant $U_2$.
This interaction corresponds to the zero-range limit of the two-channel model for the Feshbach resonance~\cite{Cui2016PhysRevA.94.043636,Tajima2021PhysRevA.104.023319}.
The relation between $U_2$ and the $p$-wave scattering length $a$ is obtained from the two-body $T$ matrix as~\cite{Cui2016PhysRevA.94.043636,Valiente2020PhysRevA.102.053304}
\begin{align}
    \frac{1}{U_2}-\sum_{p}\frac{mp^2}{k^2+i\delta-p^2}
    =\frac{m}{2}\left(\frac{1}{a}-\frac{1}{2}r_{\rm eff}k^2+ik\right),
\end{align}
where $r_{\rm eff}$ is the effective range and $\delta$ is an infinitesimally small number.
Correspondingly, we have
\begin{align}
    -\frac{1}{mU_2}=\frac{\Lambda}{\pi}-\frac{1}{2a},
\end{align}
where $\Lambda$ is the momentum cutoff.
$\Lambda$ can also be expressed in terms of the effective range $r_{\rm eff}$ as
\begin{align}
    \Lambda=\frac{4}{\pi r_{\rm eff}}.
\end{align}

For convenience, the pair-creation and -annihilation operators are introduced as
\begin{align}
    B_{k_1,k_2}^\dag = c_{k_1}^\dag c_{k_2}^\dag, \quad B_{k_3,k_4}= c_{k_4}c_{k_3},
\end{align}
respectively.
In a similar way, the corresponding operators for the three-body sector read
\begin{align}
    F_{k_1,k_2,k_3}^\dag = c_{k_1}^\dag c_{k_2}^\dag c_{k_3}^\dag, \quad F_{p_1,p_2,p_3} = c_{p_3}c_{p_2}c_{p_1}.
\end{align}

\section{Cooper-pair problem}\label{sec:III}

In this section, we first solve the Cooper pair problem in the one-dimensional $p$-wave system as described in Sec.~\ref{sec:II}. 
Correspondingly, the trial wave function is adopted as
\begin{align}
\label{eq:9}
    |\Psi_2\rangle=\sum_{k}\theta(|k|-k_{\rm F})\Phi_{k}B_{k,-k}^\dag|{\rm FS}\rangle,
\end{align}
where $|{\rm FS}\rangle$ denotes the Fermi sea.
By minimizing the ground-state energy based on the variational principle, the variational parameter $\Phi_k$ in Eq.~(\ref{eq:9}) can be determined.
In what follows, we introduce the momentum summation restricted by the Fermi surface as
\begin{align}
    \sum_{k_1,k_2,\cdots}'F(k_1,k_2,\cdots)&=\sum_{k_1,k_2,\cdots}\theta(|k_1|-k_{\rm F})\theta(|k_2|-k_{\rm F})\cdots \cr
    &\quad \times F(k_1,k_2,\cdots),
\end{align}
for an arbitrary function $F(k_1,k_2,\cdots)$, where
$k_{\rm F}=\sqrt{2mE_{\rm F}}$ is the Fermi momentum.

The expectation values for the kinetic and interaction parts are obtained as
\begin{align}
    \langle\Psi_2 \left|K\right|\Psi_2\rangle &= \sum_{k,p,q}'\xi_{p}\Phi_{k}^*\Phi_{q}\langle{\rm FS}|B_{k,-k}c_{p}^\dag c_{p} B_{q,-q}^\dag |{\rm FS}\rangle\cr
    &=2\sum_{k}'(\xi_{k}+\xi_{-k})|\Phi_k|^2
\end{align}
and
\begin{widetext}
\begin{align}
    \langle\Psi_2 \left|V\right|\Psi_2 \rangle=\,&\frac{U_2}{2}\sum_{p,q,k_1,k_2,k_1',k_2'}'
    \left(\frac{k_1-k_2}{2}\right)\left(\frac{k_1'-k_2'}{2}\right)
    \Phi_{p}^*\Phi_{q}
    \langle{\rm FS}|B_{k,-k}B_{k_1,k_2}^\dag B_{k_1',k_2'} B_{q,-q}^\dag |{\rm FS}\rangle\delta_{k_1+k_2,k_1'+k_2'}\nonumber\\
    =\,&2U_2\sum_{p,q}'
    pq
    \Phi_{p}^*\Phi_{q},
\end{align}
\end{widetext}
respectively.

Furthermore, from the variational principle, we obtain
\begin{align}
\frac{\delta\langle\Psi_2|\red{(H-E_2)}|\Psi_2\rangle}{\delta\Phi_{p}^*}=0,
\end{align}
where \red{$E_2$ is the ground-state energy of a pairing state.}
Consequently, the variational equation reads
\begin{align}
\label{eq:var}
    2(\xi_{p}+\xi_{-p}-E_2)\Phi_{p}+2U_2p\sum_{q}'q\Phi_{q}=0.
\end{align}
In order to simplify the further derivations, we introduce
\begin{align}
\label{eq:phi}
    A=\sum_{q}'q\Phi_q.
\end{align}
Substituting Eq.~(\ref{eq:phi}) into Eq.~(\ref{eq:var}), we obtain
\begin{align}
    \Phi_p=\frac{-U_2pA}{2\xi_p-E_2}.
\end{align}
Equation~(\ref{eq:phi}) can be then rewritten as
\begin{align}
    A=\sum_{p}'p\Phi_p=-U_2A\sum_{p}'\frac{p^2}{2\xi_p-E_2}.
\end{align}
Consequently, we get the bound-state equation for the Cooper pair:
\begin{align}\label{cooperpair}
    1+U_2\sum_{p}'\frac{p^2}{2\xi_p-E_2}=0.
\end{align}
By taking the momentum cutoff $\Lambda$, the two-body equation is then given as
\begin{widetext}
\begin{align}\label{2body}
    -\frac{1}{U_2}
    =\,&\int_{k_{\rm F}\leq|p|\leq \Lambda}\frac{dp}{2\pi}\frac{mp^2}{p^2-m(2E_{\rm F}+E_2)}\nonumber\\
        =\,&\frac{m(\Lambda-k_{\rm F})}{\pi}
        +\frac{m\sqrt{m(2E_{\rm F}+E_2)}}{2\pi}
        \left[\ln\left(\frac{\Lambda-\sqrt{m(2E_{\rm F}+E_2)}}{\Lambda+\sqrt{m(2E_{\rm F}+E_2)}}\right)-\ln\left(\frac{k_{\rm F}-\sqrt{m(2E_{\rm F}+E_2)}}{k_{\rm F}+\sqrt{m(2E_{\rm F}+E_2)}}\right)\right]\nonumber\\
        =\,&I_2(p_2=0,E_2-E_{\rm F}),
\end{align}
where we introduced
\begin{align}
    I_2(p_2,E)=\frac{m}{2\pi}\int dp_1\frac{\theta(|p_1+p_2|-k_{\rm F})\theta(|p_1|-k_{\rm F})}{p_1^2+p_2^2+p_1p_2-m(3E_{\rm F}+E_2)}(p_1+p_2/2)^2.
\end{align}
\end{widetext}

\begin{figure}
  \includegraphics[width=0.45\textwidth]{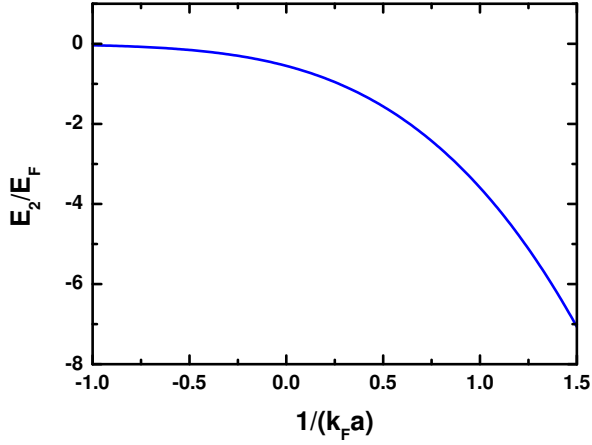}
  \caption{ 
  In-medium two-body energy 
  $E_2$ as a function of $1/(k_{\rm F}a)$ by solving the Cooper-pair problem~\eqref{2body}.
  The momentum cutoff $\Lambda$ is taken to be $10k_{\rm F}$.
  }\label{fig:1}
\end{figure}

By taking the momentum cutoff $\Lambda=10k_{\rm F}$, the ground state of the Cooper-pair problem can be solved from Eq.~\eqref{2body}.
The two-body ground-state energy $E_2$ as a function of $1/(k_{\rm F}a)$ is shown in Fig.~\ref{fig:1}.
It can be seen that the $p$-wave Cooper pair exists even on the weak-coupling side and undergoes the crossover towards the molecular state like the BCS-BEC crossover in a three-dimensional Fermi gas~\cite{Strinati2018Phys.Rep.738.1--76,Ohashi2020Prog.Part.Nucl.Phys.111.103739}.
Note that the result is qualitatively unchanged by the difference of $\Lambda$.

Here we can further see the asymptotic behavior of the two-body ground-state energy in the Cooper-pair problem.
On the one hand, in the weak-coupling limit where $|E_2|\ll E_{\rm F}$, Eq.~(\ref{2body}) further reduces to
\begin{align}
    &-\frac{1}{U_2}\simeq\frac{m(\Lambda-k_{\rm F})}{\pi}+\frac{m\sqrt{2m(2E_{\rm F}+E_2)}}{2\pi}\ln\left(\frac{|E_2|}{8E_{\rm F}}\right),
\end{align}
which indicates that
\begin{align}
    |E_2|\simeq 8E_{\rm F}e^{\frac{\pi}{k_{\rm F}a}}.
\end{align}
This expression is identical to the three-dimensional $s$-wave case after replacing the $p$-wave scattering length $a$ with the $s$-wave one~\cite{Ohashi2020Prog.Part.Nucl.Phys.111.103739},
indicating the similarity between the one-dimensional $p$-wave pairing and the three-dimensional $s$-wave one~\cite{Tajima2021PhysRevA.104.023319}.
On the other hand, in the strong-coupling limit $|E_2|\gg E_{\rm F}$, we obtain
\begin{align}
    -\frac{1}{U_2}\simeq&
    \frac{m(\Lambda-k_{\rm F})}{\pi}
    +\frac{m\sqrt{m(|E_2|-2E_{\rm F})}}{\pi}\cr
    &\times\tan^{-1}\left(\frac{\sqrt{m(|E_2|-2E_{\rm F})}}{\Lambda}\right),
\end{align}
leading to $|E_2|\simeq 2E_{\rm F}+\frac{1}{ma^2}$ in the limit of $\Lambda/k_{\rm F}\rightarrow\infty$.
This is equivalent to the two-body binding energy except for the shift $2E_{\rm F}$ associated with the Fermi sea. 

In addition, we can define the pair-correlation length~\cite{Pistolesi1994PhysRevB.49.6356,Pistolesi1996PhysRevB.53.15168} as
\begin{align}\label{pairsize}
    \xi_{\rm pair}^2=\frac{\sum_{p}'|\partial_p\Phi_{p}|^2}{\sum_{p}'|\Phi_{p}|^2},
\end{align}
where in detail, the summations read
\begin{align}
    \sum_{p}'|\Phi_p|^2
    &=2m^2U_2^2A^2
    \int_{k_{\rm F}}^{\Lambda}dp
\frac{p^2}{(p^2-k_{\rm F}^2+m|E_2|)^2}
\end{align}
and
\begin{align}
    \sum_{p}'|\partial_p\Phi_p|^2
    &=2mU_2^2A^2
    \int_{k_{\rm F}}^{\Lambda}dp
    \left[\frac{p^2+k_{\rm F}^2-m|E_2|}{(p^2-k_{\rm F}^2+m|E_2|)^2}
    \right]^2.
\end{align}

\begin{figure}
  \includegraphics[width=0.45\textwidth]{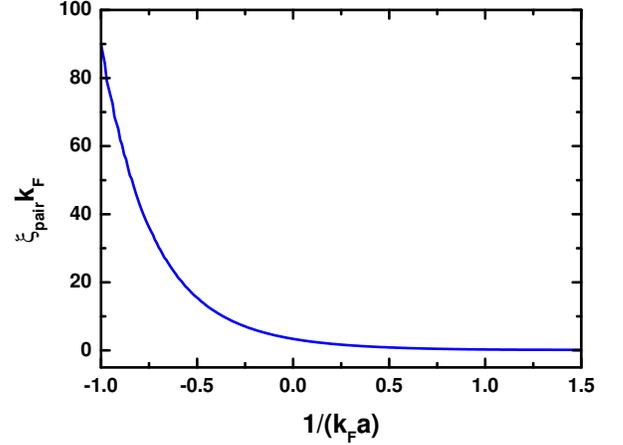}
  \caption{ 
  Pair-correlation length $\xi_{\rm pair}$ 
  as a function of $1/(k_{\rm F}a)$.
  The momentum cutoff $\Lambda$ is taken to be $10k_{\rm F}$.
  }\label{fig:4}
\end{figure}

The pair-correlation length $\xi_{\rm pair}$ given by Eq.~\eqref{pairsize},
which describes the size of the pair, is shown in Fig.~\ref{fig:4} as a function of $1/(k_{\rm F}a)$.
The momentum cutoff $\Lambda$ is taken to be $10k_{\rm F}$.
The large pair-correlation length $\xi_{\rm pair}$ on the weak-coupling side indicates the huge overlaps among pairs when a macroscopic number of Cooper pairs are formed. 
Qualitatively, the mean interparticle distance is given by $k_{\rm F}^{-1}$, and hence, $\xi_{\rm pair}k_{\rm F}\gesim 1$ represents such an overlap of pairs.
With the increase of coupling strength, the pair-correlation length $\xi_{\rm pair}$ decreases quickly.
Finally, $\xi_{\rm pair}$ becomes very small on the strong-coupling side, which indicates the formation of a molecule-like bound state.
Such behavior of the pair-correlation length is similar to the $s$-wave case in Refs.~\cite{Pistolesi1994PhysRevB.49.6356,Pistolesi1996PhysRevB.53.15168}.

Incidentally, although the Cooper problem is investigated here, the pairing energy $E_2$ is qualitatively similar to the reduction of the chemical potential $\mu$ with the pairing gap $D$ in the mean-field theory, which can be expressed as $\Delta E=2(\mu-mD^2)-2E_{\rm F}$, as discussed in the Appendix.
We note that the Cooper problem gives an approximated two-body ground state in the medium, which differs from
an exact many-body ground state.
Therefore, generally, the Cooper problem does not satisfy the Bose-Fermi duality.
However, it is still useful to understand the pairing effect in this system.
Indeed, this approach can describe both the Cooper-pair formation in the weak-coupling limit and the molecule formation in the strong-coupling limit.

\section{Three-body problem in the medium}\label{sec:IV}

In a way similar to the two-body case, the trial wave function for the three-body sector is taken to be~\cite{Kirk2017Phys.Rev.A96.053614,Tajima2021Phys.Rev.A104.053328}
\begin{align}
    |\Psi_{3}\rangle=\sum_{p_1,p_2,p_3}'\delta_{p_1+p_2,-p_3}\Omega_{p_1,p_2}F_{p_1,p_2,p_3}^\dag |{\rm FS}\rangle,
\end{align}
where $\Omega_{p_1,p_2}$ is the variational parameter and the three-body state with zero center-of-mass momentum ($p_1+p_2+p_3=0$) is considered here.
\begin{widetext}
The expectation value of the kinetic part is then obtained as
\begin{align}
    \langle \Psi_{3}\left| K\right|\Psi_{3} \rangle
    =\,&\sum_{p_1,p_2,p_3,p_1',p_2'.p_3'}'\Omega_{p_1,p_2}^*\Omega_{p_1',p_2'}(\xi_{p_1}+\xi_{p_2}+\xi_{p_3})\epsilon_{p_1,p_2,p_3}\epsilon_{p_1',p_2',p_3'}\delta_{p'_3,-p'_1-p'_2}\delta_{p_3,-p_1-p_2}\cr
    =\,&2\sum_{p_1,p_2}'(\xi_{p_1}+\xi_{p_2}+\xi_{-p_1-p_2})\Omega_{p_1,p_2}^*\left[\Omega_{p_1,p_2}+\Omega_{p_2,-p_1-p_2}+\Omega_{-p_1-p_2,p_1}\right].
\end{align}
In addition, the expectation value of the interaction part can also be derived as
\begin{align}
    \langle\Psi_{3}\left| V\right|\Psi_{3}\rangle =\,&\frac{U_2}{2}
    \sum_{k_1,k_2}'
    \sum_{k_1',k_2'}'
    \sum_{p_1,p_2,p_3}'
    \sum_{p_1',p_2',p_3'}'
    \left(\frac{k_1-k_2}{2}\right)\left(\frac{k_1'-k_2'}{2}\right)\Omega_{p_1,p_2}^*\Omega_{p_1',p_2'} \delta_{p_1+p_2,-p_3}\delta_{p_1'+p_2',-p_3'}\cr
    &\times\langle{\rm FS}|F_{p_1,p_2,p_3}B_{k_1,k_2}^\dag B_{k_1',k_2'}F_{p_1',p_2',p_3'}^\dag
    |{\rm FS}\rangle\nonumber\\
    \equiv\,&2v_1+v_2,
\end{align}
where in detail,
\begin{subequations}
\begin{align}
v_1=\frac{U_2}{2}\sum_{p_1,p_2,q}'\Omega_{p_1,p_2}^* 
\left[(p_1-p_2)(2q-p_1-p_2)\Omega_{-p_1-p_2,q}
+(2p_2+p_1)(2q+p_1)\Omega_{p_1,q}
+(-2p_1-p_2)(2q+p_2)\Omega_{p_2,q}
\right],
\end{align}
and
\begin{align}
v_2=\frac{U_2}{2}\sum_{p_1,p_2,q}'\Omega_{p_1,p_2}^* &\left[(p_1-p_2)(2q-p_1-p_2)\Omega_{q,-q+p_1+p_2}
+(2p_2+p_1)(2q+p_1)\Omega_{q,-q-p_1}\right.&\cr
&\left.
+(-2p_1-p_3)(2p_1+p_2)\Omega_{q,-q-p_2}
\right].
\end{align}
\end{subequations}

From the variational principle, we obtain
\begin{align}
\frac{\delta\langle\Psi_3|(H-E)|\Psi_3\rangle}{\delta\Omega_{p_1,p_2}^*}=\frac{\delta\langle\Psi_3|(K+V-E)|\Psi_3\rangle}{\delta\Omega_{p_1,p_2}^*}=0.
\end{align}
The resulting variational equation reads
\begin{align}\label{3body}
    &2(\xi_{p_1}+\xi_{p_2}+\xi_{p_3}-E)\left[\Omega_{p_1,p_2}+\Omega_{p_2,p_3}+\Omega_{p_3,p_1}\right]\cr
    =\,&-\frac{U_2}{2}\sum_{q}'\left[(p_1-p_2)(2q+p_3)(2\Omega_{p_3,q}
    +\Omega_{q,-q-p_3})\right.\nonumber\\
    &+\left.(p_2-p_3)(2q+p_1)(2\Omega_{p_1,q}+\Omega_{q,-q-p_1})
    \right.\cr
    &+\left.(p_3-p_1)(2q+p_2)(2\Omega_{p_2,q}+\Omega_{q,-q-p_2})\right].
\end{align}

By further introducing
\begin{align}
    \mathcal{A}(p_1,p_2)=\sum_{q}'(p_1-p_2)(2q+p_3)(2\Omega_{p_3,q}+\Omega_{q,-q-p_3})
    \equiv (p_1-p_2)\mathcal{B}(p_3),
\end{align}
Eq.~\eqref{3body} can be recast into
\begin{align}
\label{eq:32}
\left[1+U_2\sum_{p_1}'\frac{(p_1+p_2/2)^2}{\xi_{p_1}+\xi_{p_2}+\xi_{-p_1-p_2}-E}\right]\mathcal{B}(p_2)
    &=U_2\sum_{p_1}'\frac{(p_1+p_2/2)(p_1+2p_2)}{\xi_{p_1}+\xi_{p_2}+\xi_{-p_1-p_2}-E}\mathcal{B}(p_1),
\end{align}
which actually corresponds to the in-medium Skorniakov-Ter-Martirosian (STM) equation~\cite{Niemann2012Phys.Rev.A86.013628,Tajima2021Phys.Rev.A104.053328,Tajima2019New.J.Phys.21.073051}.
If we remove the constraint on the momentum summation associated with the Fermi sea as $\sum_{p_1}'\rightarrow\sum_{p_1}$,
we recover the usual STM equation for the three-body problem.
Equation~(\ref{eq:32}) can be further rewritten in terms of the in-medium two- and three-body $T$ matrices as~\cite{Tajima2019New.J.Phys.21.073051}
\begin{align}
\label{STM}
    T_2^{-1}(p_2,E)
    \mathcal{B}(p_2)=T_3(p_2,E),
\end{align}
where we introduce
\begin{align}
        T_2(p_2,E)=\left[\frac{1}{U_2}+I_2(p_2,E)\right]^{-1}
\end{align}
and
\begin{align}
\label{eq:35}
    T_3(p_2,E)&=\sum_{p_1}'\frac{(p_1+p_2/2)(p_1+2p_2)\theta(|p_1+p_2|-k_{\rm F})}{\xi_{p_1}+\xi_{-p_1}+\xi_{-p_1-p_2}-E}\mathcal{B}(p_1)\cr
    &=\frac{m}{2\pi}\int dp_1\frac{(p_1+p_2/2)(p_1+2p_2)\theta(|p_1|-k_{\rm F})\theta(|p_1+p_2|-k_{\rm F})}{p_1^2+p_2^2+p_1p_2-m(3E_{\rm F}+E)}\mathcal{B}(p_1)\cr
    &\red{\equiv -\frac{m}{4\pi}\int dp_1
    \theta(|p_1|-k_{\rm F})\theta(|p_1+p_2|-k_{\rm F})t_F(p_1,p_2)
    \mathcal{B}(p_1).}
\end{align}
\end{widetext}
In the last line of Eq.~(\ref{eq:35}),
following Ref.~\cite{Sekino2021PhysRevA.103.043307},
we have defined
\begin{align}
\label{eq:36}
    t_F(p_1,p_2)&=-\frac{2p_1^2+2p_2^2+5p_1p_2}{p_1^2+p_2^2+p_1p_2-m(3E_{\rm F}+E)}\cr
    &\equiv \frac{3p_1p_2+2m(3E_{\rm F}+E)}{m(3E_{\rm F}+E)-(p_1^2+p_2^2+p_1p_2)}
    -2.
\end{align}
One can find that Eq.~(\ref{eq:35})
exhibits an ultraviolet divergence due to the second term of $t_F(p_1,p_2)$ (i.e., $-2$) in Eq.~(\ref{eq:36}).
To avoid this ultraviolet divergence and keep the universal Bose-Fermi duality in the sense that the interaction is characterized by only $a$ while taking the large-$\Lambda$ limit,
a dimensionless three-body coupling $v_3=2$ was introduced in Ref.~\cite{Sekino2021PhysRevA.103.043307}.
Because we are interested in in-medium three-body properties with the two-body $p$-wave interaction rather than such universal properties, 
we do not go into details about $v_3$ and use the present interaction with finite $\Lambda$ in this paper.
Such a scheme is also related to the finite-range two-body interaction as given by $r_{\rm eff}=\frac{4}{\pi \Lambda}$.

By solving Eq.~\eqref{STM}, we numerically find that there is only the trivial solution for $\mathcal{B}(p)$, i.e.,
$\mathcal{B}(p)=0$, except for $E=E_{\rm sol.}$ corresponding to the continuum of in-medium pairing states and an additional fermion on the Fermi sea. 
Such a result demonstrates that the three-body bound state is absent in one-dimensional spinless $p$-wave fermionic systems with attractive two-body interaction.
As we will show below, the fermion-dimer repulsion always exists at the solution of the in-medium three-body equation $E=E_{\rm sol.}$.
It is equivalent to the solution of the following equation:
\begin{align}
\label{eq:t2}
    T_2^{-1}(p=k_{\rm F},E=E_{\rm sol.})=0
\end{align}
because the divergence of $T_2(p,E)$ in the range of $k_{\rm F}\leq p\leq \Lambda$ should be avoided in the in-medium three-body equation.

\begin{figure}
  \includegraphics[width=0.45\textwidth]{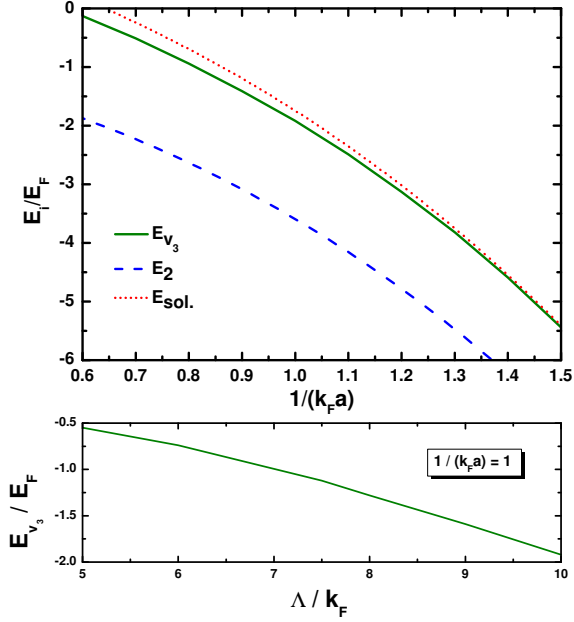}
  \caption{ 
  Top: The solution $E_{\rm sol.}$ of the in-medium three-body equation~\eqref{STM} and the Cooper-pair energy $E_2$ as functions of $1/(k_Fa)$ at $\Lambda/k_{\rm F}=10$.
  The results are shown by the red dotted and blue dashed lines, respectively.
  $E_{v_3}$, shown by the olive solid line, is the solution of the in-medium three-body bound state where we replace $t_F(p_1,p_2)$ with $t_F(p_1,p_2)+v_3\equiv t_F(p_1,p_2)+2$ in Eq.~(\ref{eq:35}) based on Ref.~\cite{Sekino2021PhysRevA.103.043307}.
  Bottom: The cutoff dependence of $E_{v_3}$ at $1/(k_{\rm F}a)=1$.
   }\label{fig:2}
\end{figure}

In order to investigate the physics deeper, we calculate $E_{\rm sol.}$ numerically.
In Eq.~(\ref{eq:t2}), $E_{\rm sol.}$ \red{can also be regarded} as the energy of the pairing state with center-of-mass momentum $k_{\rm F}$.
This configuration also appears in the in-medium three-body equation (\ref{STM}) within the zero center-of-mass frame of three particles above the Fermi sea.
In this regard, the solution of Eq.~(\ref{eq:t2}) can also be that of the in-medium three-body equation with $\mathcal{B}(p)\neq 0$.
Otherwise, the in-medium three-body equation exhibits a singularity associated with the divergent two-body $T$-matrix at an arbitrary momentum $p> k_{\rm F}$. 
In Fig.~\ref{fig:2}, the two-body ground-state energy $E_2$ obtained with the Cooper-pair problem and $E_{\rm sol.}$ are shown as functions of interaction strength $1/(k_{\rm F}a)$.
The results are shown by blue dashed and red solid lines, respectively.
The difference between $E_{\rm sol.}$ and two-body state energy $E'_2$ in the three-body problem can be estimated as
\begin{align}
    E'_2+\frac{k_{\rm F}^2}{4m}+E_{\rm F}=E_{\rm sol.}+3E_{\rm F},
\end{align}
i.e.,
\begin{align}
\label{eq:38}
    E_{\rm sol.}-E'_2=\frac{3}{2}E_{\rm F},
\end{align}
where we assume that two of the three particles on the Fermi sea have formed a Cooper pair with energy $E_2'$ and nonzero center-of-mass momentum $k_{\rm F}$, such that the total momentum with the unpaired fermion is zero.

Here the two-body state energy $E_2$ obtained with the Cooper-pair problem as shown in Fig.~\ref{fig:1} should be slightly different from $E'_2$
because $E_2$ is obtained with the in-medium two-body problem without concern for correlations with the additional third particle.
However, it can be seen that the difference between $E_2$ and $E_{\rm sol.}$ is still around $\frac{3}{2}E_{\rm F}$ in Fig.~\ref{fig:2}, as found in Eq.~(\ref{eq:38}).
This result indirectly shows that $E'_2$ is close to $E_2$ in this regime.

\begin{figure}
  \includegraphics[width=0.45\textwidth]{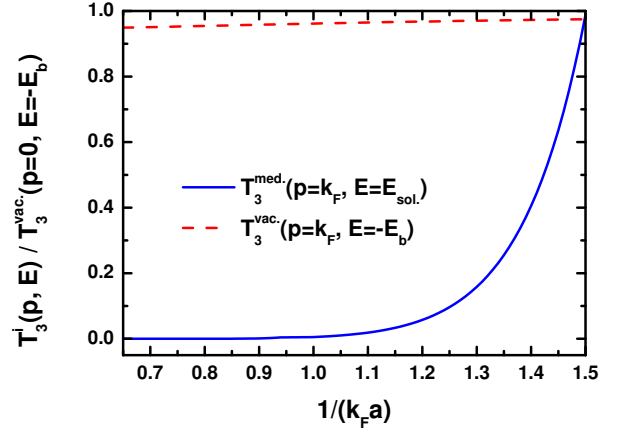}
  \caption{ 
  $T_3^{i}(p=k_{\rm F}, E=E_{\rm sol.})/T_3^{\rm vac.}(p=0, E=-E_b)$ as a function of $1/(k_{\rm F}a)$ (where ${i}={\rm med.}$ or ${\rm vac.}$).
  The results for $T_3^{\rm med.}(p=k_{\rm F}, E=E_{\rm sol.})$ and $T_3^{\rm vac.}(p=k_{\rm F}, E=-E_b)$ are shown by the blue solid and red dashed lines, respectively.
  Here, $\Lambda/k_{\rm F}=10$.
  }\label{fig:3}
\end{figure}

As pointed out in Ref.~\cite{Sekino2021PhysRevA.103.043307}, Eq.~(\ref{eq:35}) explicitly depends on $\Lambda$ because of the dimensional transmutation, and the dimensionless three-body coupling $v_3=2$ is introduced by replacing $t_{\rm F}(p_1,p_2)$ with $t_{\rm F}(p_1,p_2)+v_3$ in Ref.~\cite{Sekino2021PhysRevA.103.043307}.
In Fig.~\ref{fig:2}, we also plot the in-medium three-body binding energy $E_{v_3}$ by using this procedure.
This three-body solution at $E_{v_3}\gesim -3E_{\rm F}$ is similar to the squeezed Cooper triple discussed in Refs.~\cite{Tajima2021Phys.Rev.A104.053328,Tajima2022Phys.Rev.Research4.L012021}.
While $E_{v_3}$ is close to $E_{\rm sol.}$ on the strong-coupling side, $E_{v_3}$ gradually deviates from $E_{\rm sol.}$ at weaker coupling, indicating the sensitivity to the residual three-body interaction on the weak-coupling side. 
$E_{v_3}$ shows a strong cutoff dependence, as shown in the bottom panel of Fig.~\ref{fig:2}, where $1/(k_{\rm F}a)=1$ is adopted, while the in-vacuum three-body binding energy $4/(ma^2)$ at $\Lambda\rightarrow\infty$ was reported in Ref.~\cite{Sekino2021PhysRevA.103.043307}.
Although we specifically focus on the case with only two-body interaction and finite $\Lambda$, 
the properties of the in-medium three-body state induced by the three-body interaction require further detailed investigation, which is out of the scope of this paper.

To investigate the physics behind the absence of in-medium trimers and Cooper triples in this system \red{without the three-body interaction}, we calculate the in-medium three-body $T$ matrix $T_3(p=k_{\rm F},E=E_{\rm sol.})$ at $p=k_{\rm F}$ and $E=E_{\rm sol.}$, representing the interaction between a bound dimer and a fermion on the Fermi surface.
While our calculation does not include the lowest-order inhomogeneous term compared to the exact diagrammatic approach~\cite{Brodsky2006Phys.Rev.A73.032724},
the effective repulsive interaction given by $T_3(p=k_{\rm F},E=E_{\rm sol.})$
is sufficient for a qualitative description of in-medium fermion-dimer correlations in the strongly interacting regime.
For comparison, we calculate the in-vacuum counterpart without a Fermi sea.
Correspondingly, the energy $E$ in the two- and three-body $T$-matrices is directly fixed as the two-body binding energy, which reads
\begin{align}
E=-E_{\rm b}=-\frac{1}{ma^2}.
\end{align}
in the large-$\Lambda$ limit. 
In this paper, we employ the numerical value of $E_{\rm b}$ with $\Lambda=10k_{\rm F}$.
Moreover, for convenience, we introduce the notation of the three-body $T$ matrix as $T_{3}^{i}(p,E)$ with ${i}={\rm vac.}$ $({\rm med.})$ for the in-vacuum (in-medium) case.

In Fig.~\ref{fig:3}, we plot the ratio $T_3^{\rm med.}(p=k_{\rm F}, E=E_{\rm sol.})/T_3^{\rm vac.}(p=0, E=-E_{\rm b})$ (blue solid line) at $\Lambda/k_{\rm F}=10$.
Although it can be seen that after introducing the in-medium effect, the three-body $T$ matrix becomes smaller with the decrease of coupling strength, the interaction between the fermion and dimer is always repulsive.
To see the finite-momentum effect ($p=k_{\rm F}$) due to the presence of the Fermi sea [note that $p\geq k_{\rm F}$ in $T_{3}^{\rm med}(p,E)$], we also plot the in-vacuum counterpart $T_3^{\rm vac.}(p=k_{\rm F}, E=-E_{\rm b})/T_3^{\rm vac.}(p=0, E=-E_{\rm b})$ with $p=k_{\rm F}$ (red dashed line).
Since this in-vacuum ratio is close to $1$ in Fig.~\ref{fig:3}, the momentum dependence of the fermion-dimer scattering in vacuum is found to be weak in this parameter regime.
In this regard, the suppressed fermion-dimer repulsion in the medium is regarded as an aspect of the Fermi-surface effect.
Consequently, the three-body bound state can not be formed due to the fermion-dimer repulsion in the present system even with the Fermi sea.
Physically, while the zero-center-of-mass three-body state with $C_3$ symmetry in the momentum space near the Fermi surface (i.e., Cooper triple) can be realized in two and three dimensions~\cite{Niemann2012Phys.Rev.A86.013628,Kirk2017Phys.Rev.A96.053614,Tajima2021Phys.Rev.A104.053328,Akagami2021Phys.Rev.A104.L041302}, that is not the case in the one-dimensional geometry,
where at least one of three fermions should have a momentum away from $k_{\rm F}$.
However, it is interesting to see such a decrease of the fermion-dimer repulsion in the intermediate-coupling regime, implying the possibility of an in-medium three-body bound state with zero or nonzero center-of-mass momenta in the presence of even a small three-body attraction~\cite{Drut2018PhysRevLett.120.243002,Tajima2022Phys.Rev.Research4.L012021}.
Indeed, such a decrease in the repulsion is consistent with the qualitative behavior of $E_{v_3}$ in Fig.~\ref{fig:3}.

We note that the calculation of $E_{\rm sol.}$ is
stopped at $1/(k_{\rm F}a)\simeq 0.65$, where $E_{\rm sol.}$ becomes zero because the Cooper-pair energy and the kinetic energies of the Cooper pair and the unpaired fermion are equal to each other.
This implies that the in-medium three-body state with the zero center-of-mass momentum is not stable against the state with three non-interacting fermions on the Fermi sea, as shown in Eq.~(\ref{eq:38}).
On the other hand, as shown in the Appendix,
the strong-coupling region shown in Figs.~\ref{fig:2} and~\ref{fig:3} covers the critical value for the topological phase transition [$1/(k_{\rm F}a)\simeq 1.27$] 
predicted by the mean-field theory~\cite{Tajima2022PhysRevB.105.064508}.
Although the present variational approach cannot directly address the topological phase transition,
our result indicates the validity of the BCS-type Cooper pairing without considering the three-body clustering in such a regime.  

\begin{figure}
  \includegraphics[width=0.45\textwidth]{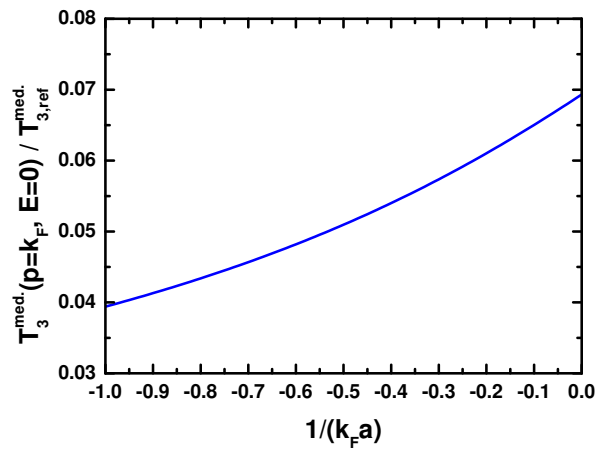}
  \caption{ 
  $T_3^{\rm med.}(p=k_{\rm F}, E=0)/T_{\rm 3,ref.}^{\rm med.}$ as a function of $1/(k_{\rm F}a)$ on the BCS side ($a<0$).
  $T_{3,{\rm ref.}}^{\rm med.}\equiv T_3^{\rm med.}(p=k_{\rm F}, E=E_{\rm sol.})$ at $1/(k_{\rm F}a)=1$ is used for a reference.
  $\Lambda/k_{\rm F}$ is also taken to be $10$.
}\label{fig:new1}
\end{figure}
Figure~\ref{fig:new1} shows the in-medium three-body $T$ matrix $T_3^{\rm med.}(p=k_{\rm F}, E=0)$ on the BCS side ($a<0$) with $\Lambda/k_{\rm F}=10$, where we used $T_{3,{\rm ref.}}^{\rm med.}\equiv T_3^{\rm med.}(p=k_{\rm F}, E=E_{\rm sol.})$ at $1/(k_{\rm F}a)=1$ as the unit instead of $T_{3}^{\rm vac.}(p=k_{\rm F},E=0)$ because this in-vacuum counterpart is extremely small in this region.
We can find that the fermion-Cooper-pair repulsion is small compared to the fermion-dimer one on the Bose-Einstein condensation (BEC) side ($a>0$) shown in Fig.~\ref{fig:3}.
This result indicates again that the BCS side may be more sensitive to the residual three-body interaction than the BEC side.

While we have shown the numerical results with mainly $\Lambda/k_{\rm F}=10$, our results for the fermion-dimer repulsion are qualitatively unchanged for the different cutoffs. However, if one were to try to figure out the competition between two- and three-body clusters in the presence of the additional three-body interaction, the cutoff dependence would play a crucial role in the entire crossover region.

\section{Summary and Perspectives}\label{sec:V}

In this paper, we have investigated in-medium two- and three-body clusters in one-dimensional spinless fermions with $p$-wave interaction.
We first solved the $p$-wave Cooper-pair problem in one-dimensional spinless fermions, and calculated the two-body bound-state energy in the medium as a function of coupling strength.
The $p$-wave Cooper pair was found to be present even in the weak-coupling limit and undergoes the crossover towards the molecular state like the BCS-BEC crossover in a three-dimensional Fermi gas with $s$-wave interaction.
In addition, the pair-correlation length, which describes the size of the $p$-wave Cooper pair, was also calculated as a function of coupling strength.
While we found a large pair-correlation length implying overlapping among pairs on the weak-coupling side, the pair-correlation length decreases with the increase of coupling strength and finally indicates the formation of tightly bound molecules on the strong-coupling side.
Furthermore, the similarity between the present Cooper problem and the results of the mean-field theory are discussed in the Appendix.

As a step further, we also investigated the corresponding in-medium three-body problem.
It can be seen that only the trivial solution can be found for the in-medium three-body equation, except for the point $E=E_{\rm sol.}$, where the $p$-wave Cooper pair with a nonzero center-of-mass momentum appears.
This result is due to the existence of the fermion-dimer repulsion associated with the one-dimensional geometry.
By making a comparison with the in-vacuum three-body $T$ matrix, we have found that although such a repulsion is weakened by the medium effect in the crossover regime, the in-medium three-body $T$ matrix is always positive.
In other words, the fermion-dimer repulsion, which suppresses in-medium three-body clustering, is found to always present.
Consequently, such a one-dimensional $p$-wave fermionic system is stable against three-body clustering even in the presence of the Fermi sea.

While we showed the absence of in-medium and in-vacuum three-body bound states in the present system with two-body interaction (without three-body interaction) at zero temperature,
our conclusion may not drastically change with finite-temperature effects because the finite-temperature effect basically suppresses the medium corrections.
If some additional factors such as a three-body force exist,
the three-body bound state can be induced as pointed out in Ref.~\cite{Sekino2021PhysRevA.103.043307}.
Indeed, following Ref.~\cite{Sekino2021PhysRevA.103.043307} for the inclusion of the dimensionless three-body coupling,
we found the solution of the in-medium bound state with the binding energy $E_{v_3}$.
In such a case, the finite-temperature effect may become important in addition to the Fermi-surface effect when the temperature $T$ exceeds $E_{v_3}$.

Our results would be useful for further investigation of unconventional superconductors and superfluids.
Moreover, the decrease in the fermion-dimer repulsion in the intermediate-coupling regime, which implies the possibility of an in-medium three-body bound state with the existence of a non-negligible three-body attraction, also paves a promising way for the study of higher-order corresponding Cooper cluster states.
Also, the emergent $s$-wave interaction due to the quasi-one-dimensional geometries may play a crucial role in the formation of Cooper cluster states~\cite{jackson2022emergent}.
Furthermore, the medium effect on bound trimers in higher dimensions such as the super Efimov state~\cite{Nishida2013PhysRevLett.110.235301} would be an interesting issue to study.

\begin{acknowledgments}
The authors are grateful to Yuta Sekino, Shoichiro Tsutsui, and Takahiro M. Doi for the useful discussions.
Y.G. was supported by the RIKEN Junior Research Associate Program.
H.T. acknowledges the JSPS Grants-in-Aid for Scientific Research under Grants No.~18H05406, No.~22H01158, and No.~22K13981.

\end{acknowledgments}

\appendix

\section{Mean-field theory}\label{appendix}

Based on BCS-Leggett theory~\cite{leggett2006quantum}, we introduce the $p$-wave superfluid order parameter as
\begin{align}
    \Delta(k)=-U_{2} k \sum_{k^{\prime}} k^{\prime}\left\langle c_{-k^{\prime}} c_{k^{\prime}}\right\rangle = k D.
\end{align}
By taking an appropriate gauge transformation, $D$ can be taken as a positive real value without loss of generality, and $\Delta(k)$ becomes real valued correspondingly. 
The mean-field Hamiltonian reads
\begin{align}
H_{\rm{MF}}=\,&
\frac{1}{2}
\sum_{k} \Psi_{k}^{\dagger}
\left(\begin{array}{cc}
\xi_{k} & -\Delta(k)  \\
-\Delta(k) & -\xi_{k}
\end{array}\right) 
\Psi_{k}\cr
&-{\frac{D^{2}}{2U_2}}+\frac{1}{2}\sum_{k} \xi_{k},
\end{align}
where
$\Psi_{k}=\left(\begin{array}{ll}
c_{k} & c_{-k}^{\dagger}
\end{array}\right)^{\mathrm{T}}$ is the Nambu spinor.
The Bogoliubov transformation is introduced here as
\begin{align}
\left(\begin{array}{c}
\alpha_{k} \\
\alpha_{k}^{\dagger}
\end{array}\right)=\left(\begin{array}{l}
u_{k} c_{k}-v_{k} c_{-k}^{\dagger} \\
u_{-k} c_{k}^{\dagger}+v_{-k} c_{-k}
\end{array}\right),
\end{align}
where $u_{k}^{2}=\frac{1}{2}\left(1+\xi_{k} / E_{k}\right)$ and $v_{k}^{2}=\frac{1}{2}\left(1-\xi_{k} / E_{k}\right)$ are the BCS coherence factors. 
Using this transformation, we obtain
\begin{align}
    H_{\rm{MF}}=\frac{1}{2}\sum_{k} E_{k} \alpha_{k}^{\dagger} \alpha_{k}+E_{\mathrm{GS}},
\end{align}
with the dispersion of the Bogoliubov quasiparticle
\begin{align}
    E_{k}=\sqrt{\xi_{k}^{2}+\Delta^{2}(k)}=\sqrt{\xi_{k}^{2}+ k^{2}D^{2} }
\end{align}
and the ground-state energy
\begin{align}
    E_{\mathrm{GS}}=-{\frac{D^{2}}{2U_2}}+\frac{1}{2}\sum_{k}\left(\xi_{k}-E_{k}\right).
\end{align}
For the given scattering length $a$ and particle number $N$, $D$ and $\mu$ can be determined by self-consistently solving the following two equations~\cite{Pastukhov2020PhysRevA.102.013307}: the gap equation,
\begin{align}\label{BCSmfr}
    \frac{m}{2 a}+\sum_{k} {k}^{2}\left[\frac{1}{2 E_{k}}-\frac{1}{2 \varepsilon_{k}}\right]\equiv\frac{1}{U_2}+\sum_{k} {k}^{2}\frac{1}{2 E_{k}}=0,
\end{align}
with $\varepsilon_{k}=k^2/(2m)$, and the particle number equation,
\begin{align}
    N=-\frac{\partial E_{GS}}{\partial \mu}=\frac{1}{2}\sum_{k} \left[1-\frac{\xi_k}{E_k}\right].
\end{align}
By further introducing the dimensionless variable $x=k/k_{\rm F}$, we can rewrite them, respectively, as
\begin{align}
\label{eq:gapeq}
    \frac{\pi}{2k_{\rm F}a}+\int_0^{\tilde{\Lambda}}dx
    \left[\frac{x^2}{\sqrt{(x^2-\tilde{\mu})^2+\tilde{D}^2x^2}}-1\right]=0
\end{align}
and
\begin{align}
\label{eq:number}
    1=\frac{1}{2}\int_0^{\infty}dx\left[1-\frac{x^2-\tilde{\mu}}{\sqrt{(x^2-\tilde{\mu})^2+\tilde{D}^2x^2}}\right],
\end{align}
where we introduced $\tilde{\mu}=\mu/E_{\rm F}$, $\tilde{D}=Dk_{\rm F}/E_{\rm F}$, and $\tilde{\Lambda}=\Lambda/k_{\rm F}$.
The solutions of Eqs.~(\ref{eq:gapeq}) and (\ref{eq:number}) 
are identical to the case of spin-$1/2$ fermions with the interspin $p$-wave interaction in Refs.~\cite{Tajima2021PhysRevA.104.023319,Tajima2022PhysRevB.105.064508}, noting that the spin degrees of freedom ($s=2$) are absorbed into the number density $\rho=s\frac{k_{\rm F}}{\pi}$.

\begin{figure}[t]
  \includegraphics[width=0.45\textwidth]{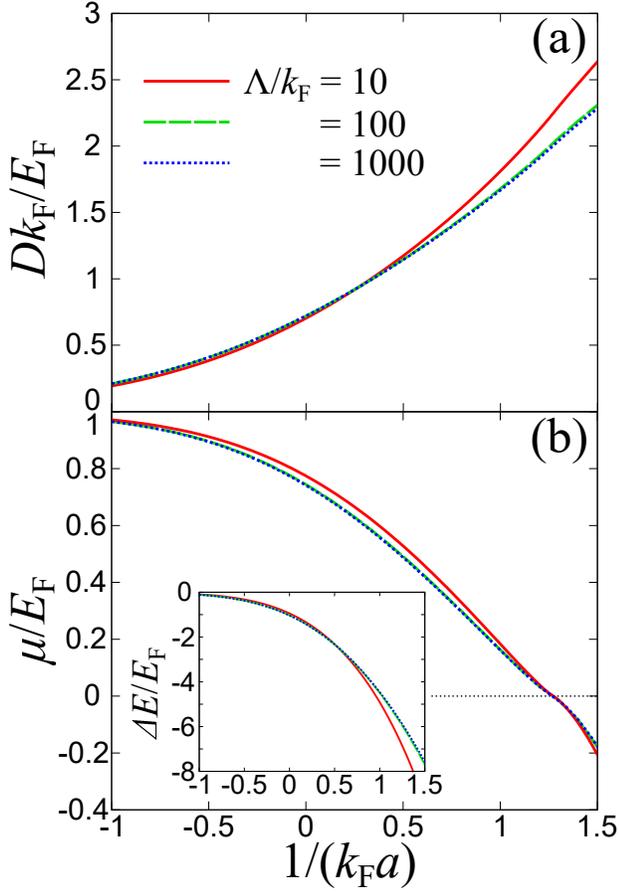}
  \caption{ 
  (a) Pairing gap parameter $Dk_{\rm F}/E_{\rm F}$ and (b) chemical potential $\mu/E_{\rm F}$ with different cutoffs $\Lambda/k_{\rm F}=10$, $100$, and $1000$ in the mean-field theory.
  The inset in (b) shows the effective two-body energy $\Delta E=2(\mu-mD^2)-2E_{\rm F}$ measured from $2E_{\rm F}$.
  }\label{fig:a1}
\end{figure}

Figure~\ref{fig:a1} shows the numerical results $Dk_{\rm F}/E_{\rm F}$ and $\mu/E_{\rm F}$ of the mean-field theory.
It is found that the cutoff dependence is not significant in the crossover regime. 
In particular, $\mu/E_{\rm E}$ is less sensitive than $Dk_{\rm F}/E_{\rm F}$,
because $\Lambda/k_{\rm F}$ is explicitly included in the gap equation~(\ref{eq:gapeq}) but not in the number-density equation~(\ref{eq:number}).
In this regard, the location of the topological phase transition ($\mu=0$)~\cite{Tajima2022PhysRevB.105.064508} is relatively robust against the change in $\Lambda$.
It is also associated with the fact that at $\mu=0$ the low-energy gapless mode $E_{k}= Dk+O(k^3)$ plays a crucial role in system's properties.
To compare the mean-field result with that in the Cooper problem in the main text,
we introduce a quantity characterizing the reduction of $\mu$ from $E_{\rm F}$ due to the pairing effect as 
\begin{align}
\Delta E=2(\mu-mD^2)-2E_{\rm F},
\end{align}
where the factor $2$ is multiplied for the comparison with the two-body energy $E_{2}$ in the Cooper problem. 
The pairing shift $mD^2$ is explicitly included in this definition because $\Delta(k)=Dk$ is not negligible even in such a relatively strong coupling regime [e.g., $1/(k_{\rm F}a)\sim 1$], in contrast to the three-dimensional $s$-wave case with the momentum-independent pairing gap~\cite{Ohashi2020Prog.Part.Nucl.Phys.111.103739}.
This fact can be found from the quasiparticle dispersion
\begin{align}
    E_{k}
    =\sqrt{\left(\frac{k^2}{2m}\right)^2-\frac{\mu-mD^2}{m}k^2+\mu^2},
\end{align}
where the term being proportional to $k^2$ involves $\mu-mD^2$.
Indeed, $\Delta E$ shown in the inset of Fig.~(\ref{fig:a1}) is 
similar to $E_2$ shown in Fig.~(\ref{fig:1}),
although $E_k$ does not exhibit the usual quadratic dispersion $\sim k^2/(2m)+|\mu|$.
In this sense, $E_2$ and $\Delta E$ does not coincide with each other quantitatively.
However, we can see that the Cooper problem and the mean-field theory give a similar result for the $p$-wave pairing in this model.
Also, we note that the mean-field theory does not satisfy the Bose-Fermi duality~\cite{girardeau1960relationship,Cheon1999PhysRevLett.82.2536,Yukalov2005,Sekino2018PhysRevA.97.013621,Valiente2020PhysRevA.102.053304,Valiente2021PhysRevA.103.L021302,Sekino2021PhysRevA.103.043307}
and the Mermin-Wagner-Hohenberg theorem~\cite{Mermin1966PhysRevLett.17.1307,Hohenberg1967PhysRev.158.383}
due to the approximation.
Nevertheless, the mean-field theory is still useful for discussing several interesting features of spinless $p$-wave fermions such as the Majorana low-energy mode~\cite{Kitaev2001}.

\end{CJK}
\end{document}